\documentclass[aps,prx,twocolumn]{revtex4-2}

\usepackage[utf8]{inputenc}
\usepackage{bm}
\usepackage{amssymb}
\usepackage{mathtools}
\usepackage{amsmath}
\usepackage{xcolor}
\usepackage{enumitem}
\usepackage{natbib}
\usepackage{hyperref}
 \hypersetup{colorlinks=true, linkcolor=blue, breaklinks=true, urlcolor=blue, citecolor=blue}
\usepackage{lipsum}
\usepackage[ruled,vlined]{algorithm2e}
\usepackage{adjustbox}

\begin{document}

\makeatletter
\newcommand*{\balancecolsandclearpage}{%
  \close@column@grid
  \clearpage
  \twocolumngrid
}
\makeatother

\title{Locating the source of forced oscillations in transmission power grids}
\author{Robin Delabays$^1$, Andrey Y. Lokhov$^2$, Melvyn Tyloo$^{2,3}$, Marc Vuffray$^2$}
\affiliation{$^1$University of Applied Sciences of Western Switzerland, Sion, Switzerland}
\affiliation{$^2$Theoretical Division, Los Alamos National Laboratory, Los Alamos, NM USA}
\affiliation{$^3$Center for Nonlinear Studies, Los Alamos National Laboratory, Los Alamos, NM USA}
\date{\today}

\begin{abstract}
        Forced oscillation event in power grids refers to a state where malfunctioning or abnormally operating equipment causes persisting periodic disturbances in the system. While power grids are designed to damp most of perturbations during standard operations, some of them can excite normal modes of the system and cause significant energy transfers across the system, creating large oscillations thousands of miles away from the source. Localization of the source of such disturbances remains an outstanding challenge due to a limited knowledge of the system parameters outside of the zone of responsibility of system operators. Here, we propose a new method for locating the source of forced oscillations which addresses this challenge by performing a simultaneous dynamic model identification using a principled maximum likelihood approach. We illustrate the validity of the algorithm on a variety of examples where forcing leads to resonance conditions in the system dynamics. Our results establish that an accurate knowledge of system parameters is not required for a successful inference of the source and frequency of a forced oscillation. We anticipate that our method will find a broader application in general dynamical systems that can be well-described by their linearized dynamics over short periods of time.
\end{abstract}

\maketitle

\section*{Introduction}

The power grid is indubitably one if not the greatest engineering achievement of the past century, as recognized by the National Academy of Engineering~\cite{constable2003century}. It is an intricate and complex network that has the size of a continent with thousands of constituents that require constant supervision. An important aspect of this surveillance is to ensure that voltage frequencies remain within a narrow band ($50\pm0.05$Hz in Europe~\cite{EN50160} and $60\pm0.05$Hz in the U.S.~\cite{kirby2003frequency}), as the violation of this requirement can cause significant damage to vital assets and results in blackouts~\cite{ALIZADEHMOUSAVI2012157}. Voltage frequency fluctuations are primarily caused by real-time imbalances between production and consumption, such as variations in the charging of electric vehicles, or a sudden gust at a wind farm. The power grid is designed to damp these random electromechanical oscillations appearing during standard operating conditions before it creates a resonance with one of the normal modes of the system. However, malfunctioning equipment or abnormal operating conditions can cause periodic disturbances that would persist over time, creating an undesirable transfer of energy across the system, an effect referred to as forced oscillations.

Whereas most forced oscillations are localized to a particular area, some may be close in frequency to one of the dominant normal modes, resulting in a system-wide response and significant energy transfers \cite{sarmadi2015inter}. Potential impacts of these wide-area oscillations include equipment failure, inadvertent tripping or control actions, and problems with the automatic generation control. This is why fast and reliable location of the source of forced oscillations is crucial in ensuring the safety and reliability of power grids. However, it remains an outstanding challenge, even when forced oscillation events are detected on the network. 
For instance, on January 11, 2019, a forced oscillation event happened across the entire Eastern Interconnection on the U.S power grid that was promptly noticed by the reliability coordinators. Nonetheless, existing tools were ineffective at identifying the source location, and a wide-area operator action did not contribute to mitigating the event \cite{NERC}. The root cause was later fortuitously identified as a faulty input from a steam turbine at a combined-cycle power plant in Florida which  forced the system to oscillate for around 18 minutes before local plant personnel removed the unit from service. The forced oscillations created by the faulty turbine had a peak-to-peak amplitude of 200 MW at the generating unit, with power swings of about 50 MW observed as far as the New England area \cite{NERC}, which shows how these disturbances have the power to affect an entire continent.
In the case of the November 29, 2005 Western American Oscillation event, a forced oscillation with amplitude 20MW originating from Alberta in Canada created a resonance effect across the entire Western Interconnection. This led to oscillations of amplitude 200 MW registered on the California-Oregon Interface, thousands of miles away from the source~\cite{sarmadi2016analysis}. This ability of forced oscillations to cause disturbances at long distances and to be amplified by the grid dynamics seriously complicates the search of their source. Forced oscillations pose a permanent threat to the power grid with more than 20 large-scale events in the past 30 years documented in the U.S., with some of them still lacking a well-identified root cause~\cite{Gho17}.

The increasing deployment of time-synchronous and distributed frequency sensors in the grid, such as Phasor Measurement Units (PMUs)~\cite{sauer2017power}, presents an opportunity for developing advanced data-driven and automated detection and localization techniques of forced oscillations. 
In the majority of cases, \emph{detection} of forced oscillations poses a limited challenge as they can be directly observed from the Fourier peaks in the signal spectrum~\cite{Zho15}. However, forced oscillations need to be differentiated from weakly damped normal modes, or free oscillations \cite{Wan16,Xie17,Ye17}. Weakly damped modes are typically analyzed through Prony analysis~\cite{Hau90}, and are mitigated via preventative measures by power system operators. On the other hand, the \emph{localization} of forced oscillations constitutes a much tougher challenge. A complete and perfect knowledge of the system and its dynamics would allow one to locate the source of a forcing~\cite{Nud13,Cab17,Del21}. However, an accurate instantaneous knowledge of the power grid dynamics appears as too strong of an assumption given that the system parameters can fluctuate on the scale of tens of minutes due to local feedback control or temperature variations~\cite{Lok18}, while the details on the system topology may be unavailable outside of the zone of responsibility for reliability coordinators.
Different methods have been proposed to circumvent this lack of information about the system. Some techniques are based on local physical properties such as the monitoring of front arrival times~\cite{Sem16a}, the evaluation of energy flow~\cite{Che13,maslennikov2017dissipating}, signal decomposition~\cite{Hua18}, or verification of the linear relation between voltage and current by multiple PMUs~\cite{Che18, Che19}. Unfortunately, these methods can be very sensitive to modeling errors such as an inaccurate assessment of the fluctuation propagation speed, or can fail to localize perturbations that are amplified by normal modes.
Black-box machine learning methods~\cite{Che00,Car04,Lee18b} have been developed with the aim of being fully model-agnostic, but suffer from a prohibitive requirement in training examples of forced oscillations events. These various shortcomings motivated recent calls from system operators and regulators to develop robust tools for performing forced oscillation analysis and localization \cite{NERC, NASPI}, which led to a further exploration~\cite{Est21,Est21b}. Nevertheless, a correct localization of the source in the case where dominant system modes are excited due to the resonance phenomenon remains an outstanding challenge.

In this paper, we propose a new principled method of detection and localization of forced oscillations which is agnostic to the knowledge of the system topology and parameters, fully capable of identifying the source of distant normal modes excitation, and which does not rely on any offline training. Our method operates within a much broader framework that extends rather universally to any dynamical system that can be well-described over short period of time by its linearized dynamics. This makes it a method of choice for the detection and localization of energy transfers created by small disturbances on a large class of complex networks.
The key insight in our method consists in leveraging random frequencies fluctuations naturally present in the system to build in real time an effective dynamic model of the network. The dynamic model identification and source localization are performed at the same time using a principled maximum likelihood approach. We illustrate the performance of our approach on a number of examples where forcing excites the natural system modes and create a resonance phenomenon, as well as on a real-world PMU data set.
\begin{figure*}[!htb]
 \centering
 \includegraphics[width=\textwidth]{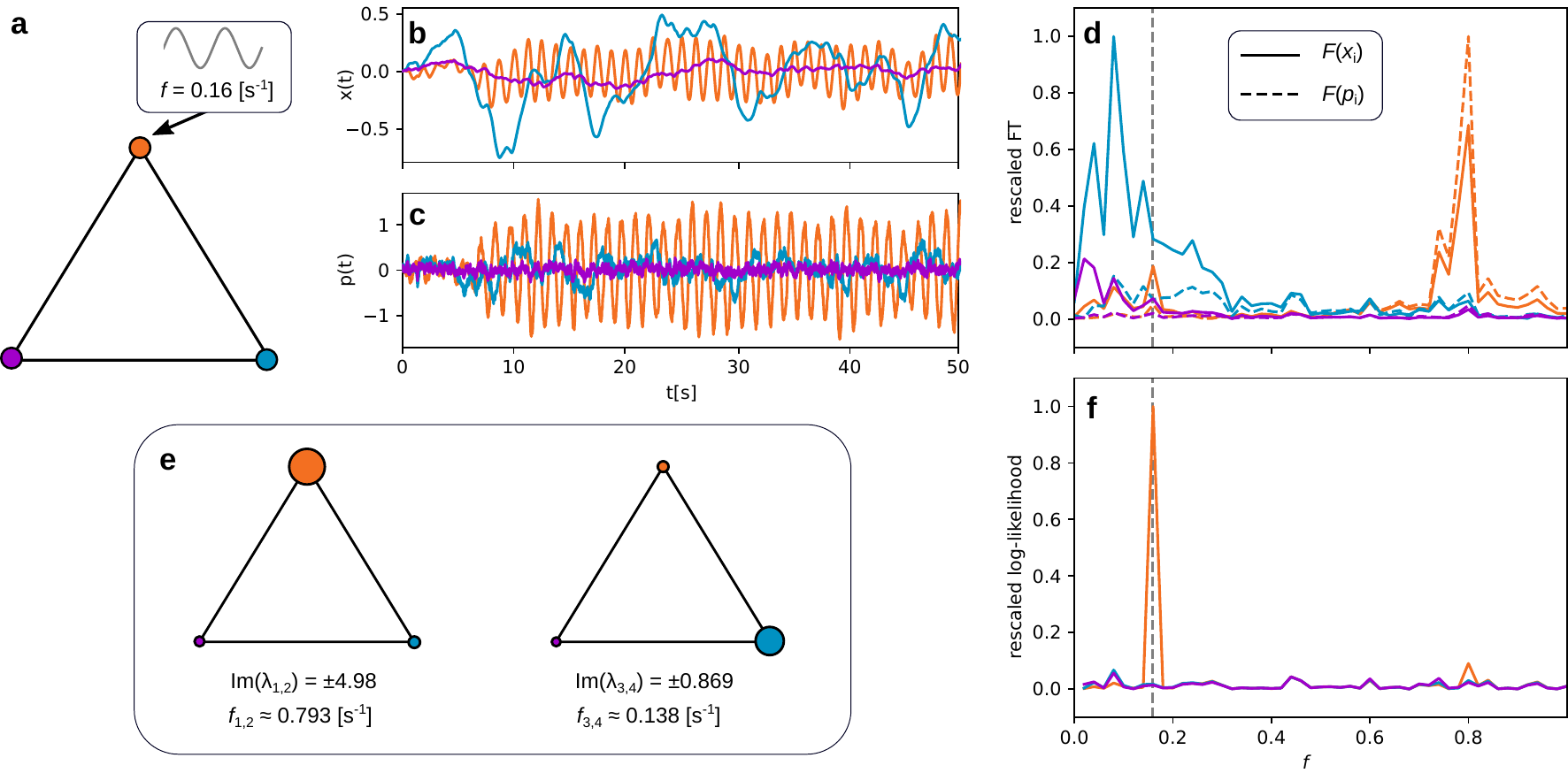}
 \caption{{\bf A toy example with three nodes illustrating the challenges and solutions for locating the source of forced oscillations in a network of coupled oscillators.} {\bf (a)} 
 A forced oscillation is induced on the orange node with frequency $f = 0.16$ [${\rm s}^{-1}$]. Time series of the network states corresponding to generalized positions $\bm{x}_t$ and momentum $\bm{p}_t$ (corresponding to phase deviations $\bm \theta_t$ and frequency deviations $\bm \omega_t$ in power grids, see \emph{Supplementary Information}, section~S1~\cite{supmat}) presented in panels {\bf (b)} and {\bf (c)}, respectively, are generated using Eq.~\eqref{eq:dynamics} and the parameters detailed in the \emph{Supplementary Information}, section S2~\cite{supmat}. {\bf (d)} A na\"ive Fourier analysis does not allow one to identify either the forcing source or location: the Fourier transform of the position time series displays its largest peak at the blue node with the frequency $f=0.08$ $[{\rm s}^{-1}]$, and the Fourier transform of the momentum time series displays peaks at the orange node, with a frequency $f=0.8$ [${\rm s}^{-1}$]. {\bf (e)} An eigendecomposition of the dynamic state matrix shows the natural frequencies (eigenvalues) and modes (eigenvectors, with the size of the nodes indicating the magnitude of the respective components) of the system. This analysis reveals the reason for the observed behavior of the Fourier transforms: the forcing frequency is close to a natural frequency of the system, thus creating a resonance effect and exciting the natural modes of the system. {\bf (f)} Our source localization algorithm confidently points to the correct source (orange node) and frequency (indicated with a dashed gray line) of the forced oscillation. The accuracy of the determination of the forcing frequency $f$ is fundamentally defined by the duration of the available time series.
 }
 \label{fig:ntw3}
\end{figure*}

\section*{Challenges and solutions for locating the oscillation source}

Learning of the dynamics of linear stochastic systems lies at the foundation of our approach. In the ambient regime of fluctuations, a complex system such as power grid with a general non-linear dynamics is typically found in a state which is close to a stable equilibrium. In this ambient regime of small perturbations, the dynamics of the system is well described by a linear stochastic equation corresponding to a celebrated model of coupled harmonic oscillators. Without surprise, this family of models includes the popular swing equations describing the ambient dynamics of generators in a transmission power grid, which are derived under the assumption of small deviations from the steady state \cite{Ber00, kundur1994power} (see \emph{Supplementary Information}, section S1~\cite{supmat} for more details).

We consider the linear stochastic network dynamics with an additional forcing at a single node $l$, indexed by $\bm{e}_l$, the canonical basis vector with nonzero $l$th component. Mathematically, this dynamics is described by the following linear stochastic differential equation,
\begin{align}
    \bm{M} d\bm{p}_t &= \bm{D} \bm{p}_t dt + \bm{L}\bm{x}_t dt + \gamma \bm{e}_l \cos(2\pi (f t + \phi)) dt + d\bm{W}_t,
    \label{eq:dynamics}
\end{align}
where $\bm{x}_t$ represents the network state variables (for a power grid, they correspond to deviations of phases from the steady state values), $\bm{p}_t dt = d \bm{x}_t$ is the generalized momentum (deviation of frequencies from the steady state values in a power grid), $\bm{M}$, $\bm{D}$, $\bm{L}$ are generalized mass, damping, and network coupling parameters, $\gamma$, $f$, and $\phi$ are the amplitude, frequency and phase of the forcing, respectively, and $\bm{W}_t$ is a Wiener noise process describing random fluctuations.

Our goal is to reconstruct the oscillation frequency and the location of the forcing source from the measured time series of $\bm{x}_t$ and $\bm{p}_t$ of length $T$ at a sequence of $N$ discrete time steps $t\in\{t_1,...,t_N\}$. We do not assume any knowledge of the system parameters or topology, which represents a realistic scenario in power grids: instantaneous awareness of system parameters are almost never available to system operators \cite{Lok18}.
Hence, we do not suppose any knowledge on the inertia $\bm{M}$, damping $\bm{D}$, or grid Laplacian matrix $\bm{L}$. Obviously, we also do not assume any knowledge of the parameters related to the forcing, i.e., $\gamma$, $l$, $f$, or $\phi$. All these parameters of interest need to be recovered from the noisy data $\{\bm{x}_t\}$ and $\{\bm{p}_t\}$.

In Figure~\ref{fig:ntw3}, we illustrate with a toy example of a three-node network the key mechanism that renders the localization of forced oscillations elusive to standard signal processing analysis. Namely, a forcing with a frequency close to a natural frequency of the system may excite the natural modes of the system peaked at a different node. As a result, neither the correct forcing frequency nor the source of the oscillations are evident from the Fourier spectrum in Figure~\ref{fig:ntw3}~(d). This effect is reminiscent of forced oscillation events in power grids such as those of November 29, 2005 and January 11, 2019 discussed above, where large perturbations can be seen far from the source and at a very different frequency.

To address this challenge, we develop a principled method for determining the oscillation frequency $f$ and locating the source of forced oscillations based on a maximum likelihood approach (see \emph{Methods} and \emph{Supplementary Information}, section S3~\cite{supmat} for a detailed derivation). A na\"ive real-space estimator leads to a complicated non-convex optimization problem in frequency, as we explain in the \emph{Supplementary Information}, section S4~\cite{supmat}. A key insight which leads to an efficient solution consists in realizing that the finite length of the time series imposes a finite resolution on the frequencies. This leads to a tractable formulation featuring the Fourier transformed quantities, which can be efficiently solved with the state-of-the-art interior-point method solvers (see \emph{Methods}).

Knowledge of the number of sources leads to a discrete formulation of the problem, where optimization is run for every node (or every pair of nodes if two sources of forcing are present, \emph{etc.}). We refer to this method as to the \emph{System Agnostic Localization of Oscillations (SALO)} algorithm. SALO approach is fully parallelizable over all nodes in the network, making it the method of choice for a high performance computing system. An example of an application of the SALO framework is given in Figure~\ref{fig:ntw3}~(e) for our toy resonance example: both the source and the frequency of forced oscillations are unambiguously and correctly identified. For streaming applications where a faster identification is desired, we consider a computationally advantageous relaxation of the problem. In this SALO-relaxed version of the algorithm, all nodes are formally allowed to be a source with a respective amplitude $\gamma_i$ for each node $i$ (see \emph{Methods} for more details). In what follows, we benchmark the method on a number of simulated and real use cases.

\section*{Tests on synthetic systems and real data}

\begin{figure*}[!htb]
 \centering
 \includegraphics[width=\textwidth]{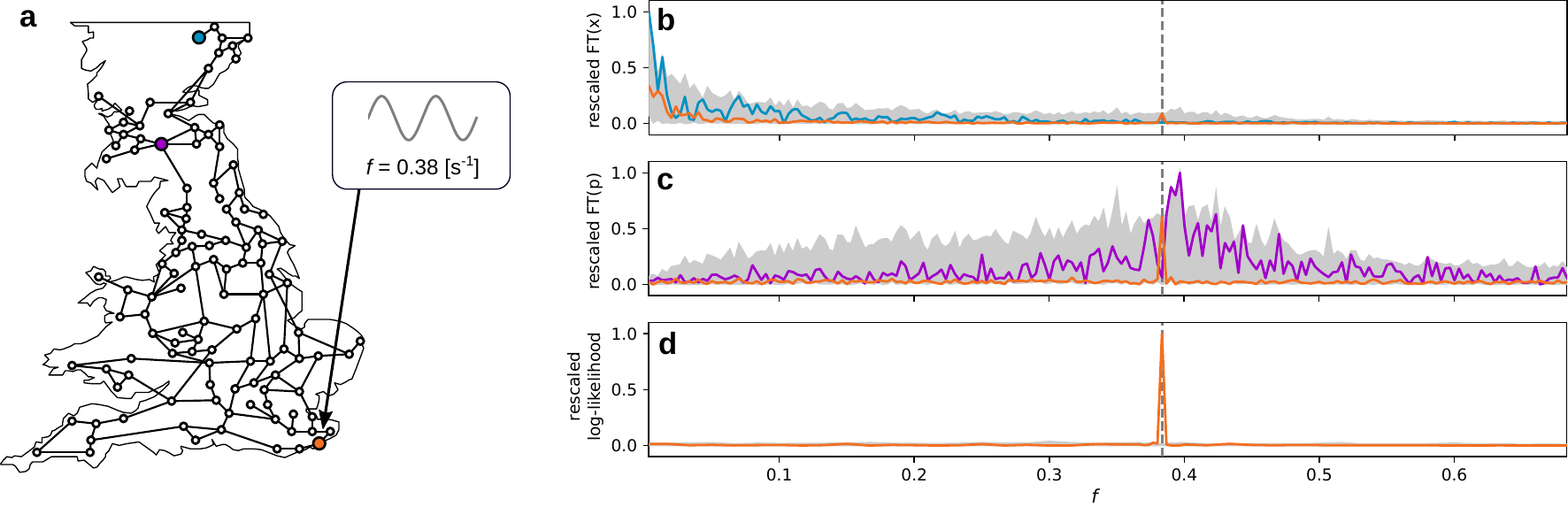}
 \caption{ {\bf Detection and localization of forced oscillations under the resonance conditions.}
 {\bf (a)} Synthetic test case with a topology inspired by the UK high-voltage grid, designed to reproduce features observed in real oscillatory events where oscillations interacted with the system modes. Details on the network parameters are given in the \emph{Supplementary Information}, section S2~\cite{supmat}. The forcing at the orange node results in a highest response in Fourier spectra at the opposite side of the network, as shown for the Fourier components of the {\bf (b)} generalized state and of the {\bf (c)} generalized momentum. {\bf (d)} SALO algorithm confidently identifies the correct forcing frequency and source without any knowledge of the system topology or parameters. The envelope of scores for non-highlighted nodes is shown in gray.
 }
 \label{fig:uk}
\end{figure*}

An effect where interaction between the forcing frequency and one of the natural modes of the system leads to a peak response away from the forcing source may arise in networks with a much more complex structure compared to the three-node system illustrated in Figure~\ref{fig:ntw3}. Such a resonance behavior represents an outstanding challenge for detection algorithms whereby the peaks in the Fourier spectrum may not only involve nodes far away from the source, but also point to frequencies related to the natural modes of the system rather than to the frequency of the forcing. We showcase this phenomenon on a synthetically generated data set according to the model \eqref{eq:dynamics} on a network inspired by the UK high-voltage grid, see Figure~\ref{fig:uk}. This test case demonstrates some of the features observed in the 2005 Western Interconnection and in the 2019 Eastern Interconnection oscillation events. In particular, the largest amplitudes in the Fourier components can be located very far away from the source, at distances comparable to the diameter of the network. On the other hand, we see that the maximum likelihood based SALO method identifies the correct forcing frequency and the correct source, for which the Fourier signals are otherwise completely hidden among the responses of other nodes in the network, see Figure~\ref{fig:uk} (b), (c), and (d).

In the \emph{Supplementary Information}, section S5~\cite{supmat}, we further illustrate the challenges of Fourier-based source localization under the resonance phenomenon on a standard IEEE test case topology with 57 nodes. Yet in this case again, as shown in the Figure~S3, SALO method precisely and unequivocally identifies the correct frequency and location of the forcing source, without exploiting any prior knowledge on the system topology and parameters. We also used this test case to demonstrate the performance of the SALO-relaxed version of the algorithm, which accurately points to the correct location of the source of forced oscillations, while benefiting from the computational complexity of a single source verification under the full maximum likelihood approach. This shows that the SALO-relaxed version can be used for a quicker assessment of the forced oscillations once they have been detected in the system, prior to running parallelized computations under the SALO framework. We further show this computational advantage on a series of synthetic instances of increasing size in Figure~S4, whereby the ratio of run-times of SALO and SALO-relaxed scales linearly with the system size.

\begin{figure*}
 \centering
 \includegraphics[width=\textwidth]{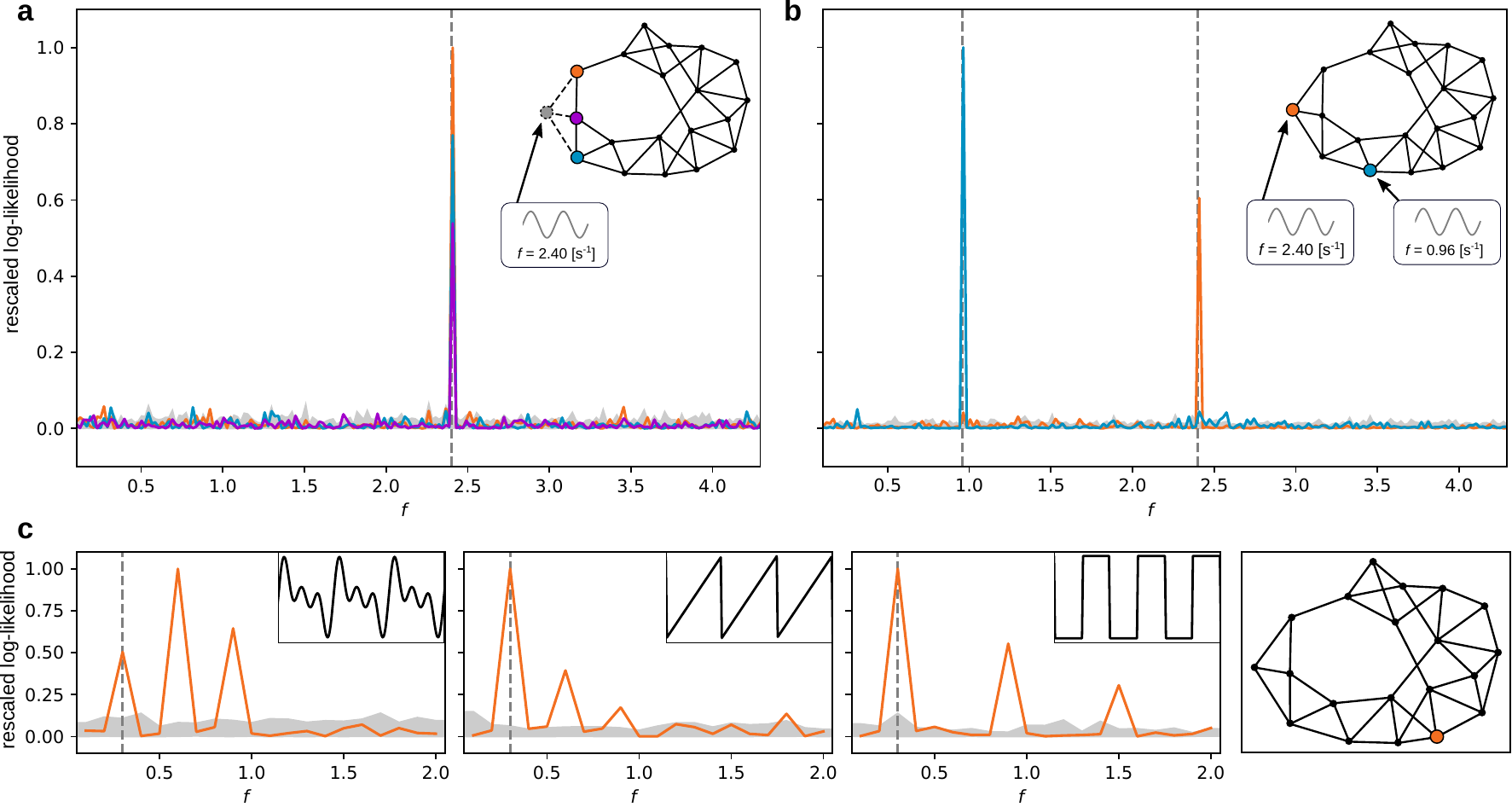}
 \caption{{\bf Robustness of the SALO algorithm to model misspecification.} {\bf (a)} In the case where the source of forced oscillations is outside of the observed system (gray node), the algorithm still correctly identifies the forcing frequency and the neighbors of the hidden source inside the visible system (three overlapping peaks in likelihood at the forcing frequency correspond to orange, purple, and blue nodes). 
 {\bf (b)} In the case of several sources, both locations appear as peaks in the rescaled likelihood score at their respective forcing frequencies. {\bf (c)} In the case of non-sinusoidal forcing injected into the system, the forcing location is still correctly identified, while the complex nature of the forcing shows up as likelihood peaks at different harmonics of the forcing signal. The envelope of scores for non-highlighted nodes is shown in gray in all panels.}
 \label{fig:fig4}
\end{figure*}

In the tests described above, we assumed that the time series have been produced using the model \eqref{eq:dynamics}, albeit the system parameters are not known to the reconstruction algorithm. In particular, it is assumed that there is a \emph{single} and \emph{observed} source node $l$, and that the forcing of the type $\cos(2\pi (f t + \phi))$ is associated with a single frequency $f$. In Figure~\ref{fig:fig4} we look at the results produced by SALO in situations where these assumptions are violated. In Figure~\ref{fig:fig4}~(a), the source node is outside of the observable system. We see that in this case, the SALO algorithm is still able to correctly identify the forcing frequency, and points to the immediate neighbors of the source node inside the observable system. The Figure~\ref{fig:fig4}~(b) demonstrates the case where two sources of oscillations at different nodes and frequencies are simultaneously present in the system. Notably, two peaks corresponding to both forcings appear in the rescaled likelihood score. Finally, in Figure~\ref{fig:fig4}~(c), we show the rescaled likelihood scores for the input forcing signals of different types. Even in this case, SALO correctly identifies the source location showing up as several peaks in the likelihood at different harmonics of the input forcing signal. This makes the algorithm remarkably robust to the assumptions behind the model. In the \emph{Supplementary Information}, section S6~\cite{supmat}, we show an example of an application of the SALO algorithm to real data which display a combination of features observed in Figure~\ref{fig:fig4}.          

\begin{figure*}[!htb]
 \centering
 \includegraphics[width=\textwidth]{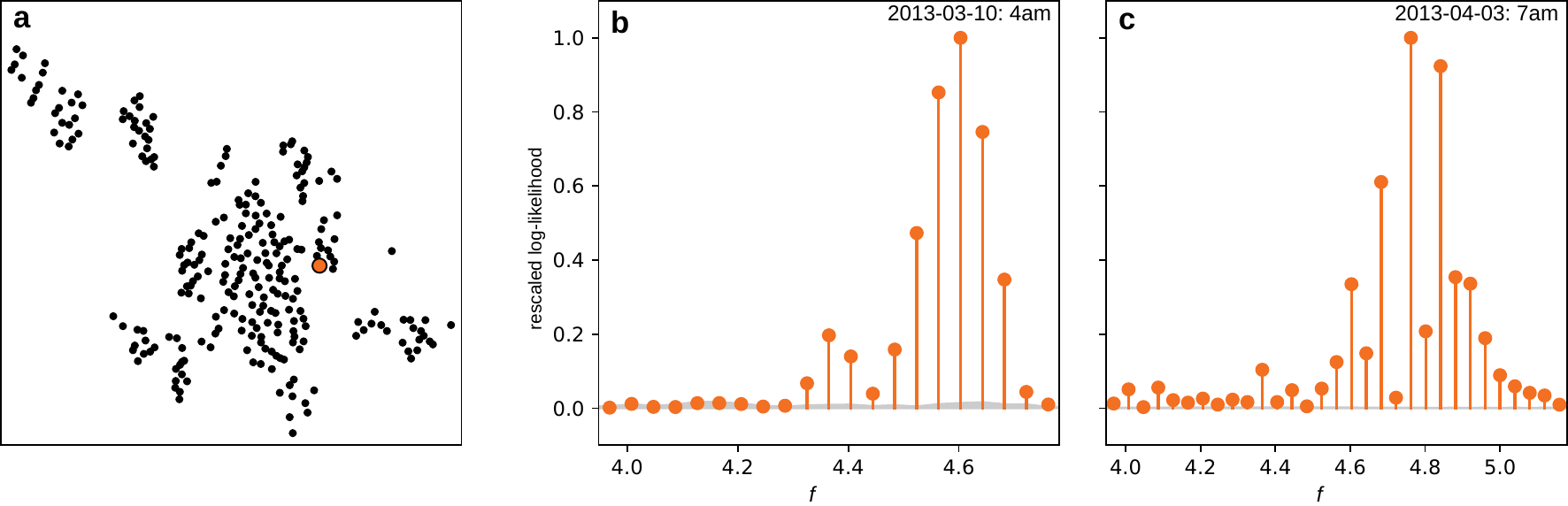}
 \caption{
 {\bf Identifying the location of forced oscillations for PMU data from a U.S. power system operator.} {\bf (a)} Geographical layout of PMU sensors in the system providing time series data (anonymized and modified coordinates). The node identified as the most likely source of forced oscillations is highlighted in orange. {\bf (b)} According to the SALO algorithm, the identified node has a much higher likelihood for being a source for a range of frequencies in the vicinity of $f=4.6$ $[{\rm s}^{-1}]$ compared to the rest of the nodes in the network (corresponding likelihoods depicted in gray). A finite range of candidate frequencies is a realistic feature that may emerge in real systems due to finite length of the collected data. {\bf (c)} The algorithm consistently points to the same candidate source in a similar frequency range even for time series collected over two time periods separated by almost a month (March 2013 vs. April 2013). This consistency strongly indicates that sustained oscillations may originate from a single faulty component of the system. The envelope of scores for other non-highlighted nodes in the system is shown in gray.
 }
 \label{fig:fig3}
\end{figure*}

Another possible misspecification is the assumption of linearity of the dynamics. For instance, real power systems do not exactly follow the linear model of the type \eqref{eq:dynamics} with constant system parameters. Instead, a linear swing model which falls within the class \eqref{eq:dynamics} as discussed in the \emph{Supplementary Information}, section S1~\cite{supmat}, represents an approximation to a general non-linear dynamics of generators under small deviations from the steady state, valid over finite periods of time \cite{Lok18, hannon2021real}. However, with the understanding that \eqref{eq:dynamics} only serves as an \emph{effective} model providing an adequate description of the complex system dynamics over short time scales, we show that the SALO method is still applicable to data from real transmission power grids with the purpose of identifying the conditions pertaining to forced oscillations. Here, we benchmark the SALO algorithm on an instance of real PMU data from a U.S. transmission power grid with 200 nodes, where a presence of sustained oscillations has been suggested in previous studies. A Fourier-type analysis of the PMU data from a U.S. Independent System Operator showed a presence of sustained oscillations with a frequency in the 4-6[s\textsuperscript{$-1$}] range, responsible for the emergence of correlations between several node clusters \cite{Esc19}. The analysis of these time series with the SALO method, see Figure~\ref{fig:fig3}, confidently points to a single source of oscillations, with a frequency close to the range previously identified in \cite{Esc19}. Incidentally, although the ground truth for this system is unknown, the identified node is consistently pointed to as the most likely source of sustained oscillations even for PMU data separated by a time interval of about a month, see Figure~\ref{fig:fig3} (b) and (c). An extended range of candidate frequencies is likely to be connected to the fundamental limits on the data resolution, as exemplified in the \emph{Supplementary Information}, section S4~\cite{supmat}, where shorter time series lead to a wider log-likelihood objective function in the frequency domain (see Figure~S1).

\section*{Conclusions}

We proposed a rigorous maximum-likelihood-based framework enabling a simultaneous system identification and localization of the source of forced oscillations at the same time, without any prior knowledge of the system topology or parameters. In particular, our method is able to perfectly locate the oscillation source, even in the case where the forcing excites one of the natural modes of the system and creates an amplitude peak at a far-away node and at a different frequency. This scenario is reminiscent of some of the real-world historical events, such as the 2005 Western Interconnection event, which turned out to be the most challenging from the oscillation source localization perspective. The ease of parallelization and a relaxed version of the algorithm makes the method scalable to large network instances and multiple sources, while robustness to modeling assumptions makes the algorithm applicable to data produced by real-world dynamical systems.

The SALO algorithm can be naturally adapted to the situation where additional prior information is available. For instance, the matrices $\bm{M}$, $\bm{D}$, and $\bm{L}$ do not need to be reconstructed from data if a prior knowledge on the network structure and parameters is available. 
In the \emph{Supplementary Information}, section~S7~\cite{supmat}, we show that under this scenario, the SALO algorithm is able to identify the correct source and frequency of the forcing using significantly less data compared to the most challenging setting considered in this work, where no prior information on the system is available.
Similarly, knowledge of the disturbance type or number of sources can also be directly incorporated into the algorithm. In the context of power grid applications, it could be practical to extend the SALO method to the case of partial sensor coverage, as well as to a source localization under the model which specifically takes into account different types of generators, buses, and PMUs in the grid. 
For instance, the primary setting considered in our work assumes that all buses have non-zero inertia, which is true for generators in the power grid, but not necessarily for other buses, e.g. those representing loads.

Due to a wider applicability of our approach to general stochastic linear dynamics and coupled harmonic oscillators, we anticipate that our methods will find a broader range of applications beyond power grids, e.g., vehicular platoon subjected to malfunctioning elements or malicious attacks, and more broadly multi-agent systems where automated units must reach an overall consensus~\cite{1272910}, as well as forced oscillations in wave propagation dynamics.

\section*{Methods}

\paragraph*{\bf Model.} The dynamic equation~\eqref{eq:dynamics} can be reformulated as a first-order linear dynamic system 
\begin{align}
    d\bm{p}_t &= \bm{A}\bm{X}_t dt + \gamma\bm{e}_l{\rm Re}\left(e^{2\pi i(f t + \phi)}\right)dt + d\bm{W}_t,
    \label{eq:dynamics_linear_formulation}
\end{align}
where we used $\bm{p}_t dt = d \bm{x}_t$, and denoted $\bm{X}_t = (\bm{x}_t, \bm{p}_t)$. Discretizing Eq.~\eqref{eq:dynamics_linear_formulation} with an Euler-Maruyama approximation scheme over $N$ points yields the finite difference equation, for $j=0,...,N-1$,
\begin{align}\label{dyn_disc}
 \bm{\Delta}_{t_j} &= \bm{A}\bm{X}_{t_j} + \gamma\bm{e}_l{\rm Re}\left(e^{2\pi i(k\frac{j}{N} + \phi)}\right) + \bm{\xi}_j\, ,
\end{align}
where $\bm{\Delta}_{t_j}=(\bm{X}_{t_{j+1}}-\bm{X}_{t_{j}})/\tau$ for $\tau=T/N$, we assumed $t_j=j\tau$, and the frequency relates to the integer $0<k<N/2$ as $k = f T$. As discussed in the problem formulation above, the measurements $\{\bm{X}_{j\tau}\}_{j=0,\ldots,N-1}$ are assumed to be available, but the parameters of the system $\bm{A}$, $\gamma$, $l$, $f$, and $\phi$ are unknown, and need to be estimated from the data.

\paragraph*{\bf Reconstruction algorithm.}

Given that on finite time intervals, the contribution of the Wiener process is an i.i.d. Gaussian variable with zero mean, i.e., $\bm{\xi}_{t_j}\sim{\cal N}(0,\tau)$, the negative log-likelihood reads,
\begin{widetext}
\begin{align}\label{eq:L_naive}
 L\left(\bm{A},\gamma,l,k,\phi~|~\{\bm{X}_{t_j}\}_{j=1}^{N}\right) &= \frac{1}{N}\sum_{j=0}^{N-1}\left\|\bm{\Delta}_{t_j} - \bm{A}\bm{X}_{t_j} - \gamma\bm{e}_l{\rm Re}\left(e^{2\pi i(\frac{k}{N}j + \phi)}\right)\right\|^2.
\end{align}
\end{widetext}
Note that a discretized set of frequencies appearing in the forcing term as a result of finiteness in $T$ and $N$ is crucial, because an optimization over a continuous variable $f$ leads to a hard nonlinear optimization problem, as exemplified in the \emph{Supplementary Information}, section S4~\cite{supmat}. For a fixed frequency $k$ and node $l$, the joint minimization over $\bm{A}$, $\gamma$, and $\phi$ remains challenging to solve as the negative log-likelihood possesses numerous local minima. Nevertheless, the global minimization of \eqref{eq:L_naive} over the phase $\phi$, as a function of the remaining parameters, can be performed exactly, leading to the following expression for the partial negative log-likelihood:
\begin{widetext}
\begin{align}\label{eq:L0_smart}
  L_{\rm SALO}\left(\bm{A},\gamma,l,k~|~\{\bm{X}_{t_j}\}_{j=0}^{N-1}\right) &= {\rm Tr}(\bm{A}^\top\bm{A}\Sigma_0) - 2{\rm Tr}(\bm{A}\Sigma_1) + \frac{1}{2}\gamma^2 - \frac{2\gamma}{\sqrt{N}} \sqrt{{\rm Tr}\left(\bm{A}_{l,\cdot}^\top \bm{A}_{l,\cdot} F(k)\right) - 2f_l(k) \bm{A}_{l,\cdot} + g_l(k)}\, ,
\end{align}
\end{widetext}
see \emph{Supplementary Information}, section S3~\cite{supmat} for details of the derivation and the definitions of all the quantities that can be directly computed from the available time series $\{\bm{X}_{j\tau}\}_{j=1,\ldots,N}$. For a fixed $l$ and $k$, the expression \eqref{eq:L0_smart} remains a non-convex function in the arguments $\bm{A}$ and $\gamma$. However, we have observed that this cost function, unlike its naive counterpart \eqref{eq:L_naive}, seems to always possess a single minimum that can be efficiently found using state-of-the-art interior point methods. In this work, we used the optimization software Ipopt \citep{wachter2006implementation} within the Julia/JuMP modeling framework for mathematical optimization \citep{DunningHuchetteLubin2017}. Note that the optimization is fully parallelizable over $k$ and $l$. We refer to the minimization of the partial negative log-likelihood as to the SALO method.

We also consider an accelerated method referred to as the SALO-relaxed method, which exploits the spatial relaxation of the problem in the variable $\gamma$. Under this variant of the algorithm, all nodes are formally allowed to be a source, and their relative likelihood to be a source is encoded by a respective amplitude $\gamma_i$ for each node $i$ (so that the regression now runs over the vector $\bm{\gamma}$):
\begin{widetext}
\begin{align}\label{eq:L1_smart}
 L_{\rm SALOr}\left(\bm{A},\bm{\gamma},k~|~\{\bm{X}_{t_j}\}_{j=1}^N\right) &= {\rm Tr}(\bm{A}^\top\bm{A}\Sigma_0) - 2{\rm Tr}(\bm{A}\Sigma_1) + \sum_{l=1}^n \left[\frac{\gamma_i^2}{2} - \frac{2\gamma_i}{\sqrt{N}}\sqrt{{\rm Tr}\left(\bm{A}_{l,\cdot}^\top\bm{A}_{l,\cdot}F(k)\right) - f_l(k)\bm{A}_{l,\cdot} + g_l(k)} \right]\,.
\end{align}
\end{widetext}
We benchmark this method in the \emph{Supplementary Information}, section S5~\cite{supmat}, where we also discuss the computational speed-up of SALO-relaxed compared to the SALO estimator.

\section*{Data availability}
All data that support the plots within this paper and other findings of this study are available from the authors on reasonable request.

\section*{Code availability}
Code implementing SALO and SALO-relaxed algorithms in Julia can be found at the following Github repository: \url{https://github.com/lanl-ansi/SALO}.

\section*{Acknowledgements}
This work was supported by U.S. DOE/OE as part of the DOE Advanced Sensor and Data Analytics Program and by the Laboratory Directed Research and Development program of Los Alamos National Laboratory under project numbers 20220797PRD2, 20210078DR, and 20220774ER. 
RD was supported by the Swiss National Science Foundation, under grant nr. P400P2\_194359. 
The authors are grateful to Prof. Daniel Bienstock from Columbia University, and Dr. Yilu Liu and Dr. Wenpeng ``Wayne'' Yu from the Power IT Lab at Oak Ridge National Lab and UT Knoxville for sharing real PMU data used in this work. The authors also thank Dr. Slava Maslennikov from ISO New England and Prof. Michael Chertkov from University of Arizona for fruitful discussions.\\

\section*{Author contributions}
All authors designed and performed the research, wrote the manuscript, reviewed and edited the paper.

\section*{Competing interests}
\noindent The authors declare no competing interests.\\

\section*{Additional information}
\noindent {\bf Supplementary information} is available for this paper~\cite{supmat}.\\
{\bf Correspondence and requests for materials} should be addressed to MV.

\balancecolsandclearpage

\newpage

\onecolumngrid

\appendix

\begin{center}
{\LARGE Supplementary information}
\end{center}

\section{Linear second-order dynamics and relation to swing equations}
\label{app:swing_equations}

In this section, we show that the linearized swing equations describing the voltage phase dynamics in the ambient regime \cite{Ber00, kundur1994power}, falls in the class of linear dynamic models [Eq. (1) in the main text] that we consider in this work. When the system remains in the vicinity of a steady-state fixed point, the linearized equations give a fair approximation of the swing dynamics~\cite[Sec.~14.3]{Ber00} at node $i\in\{1,...,n\}$:
\begin{align}\label{eq:swing}
\begin{split}
    \dot{\theta}_i &= \omega_i, \\
    m_i\dot{\omega}_i + d_i\omega_i &= - \sum_{j=1}^nb_{ij}(\theta_i-\theta_j) + \eta_i\, ,
\end{split}
\end{align}
where $\theta_i$ and $\omega_i\coloneqq \dot{\theta}_i$ are respectively phase and frequency deviations from the fixed point, $m_i$ and $d_i$ are respectively the effective inertia and damping, $b_{ij}$ is the susceptance of the line between nodes $i$ and $j$ (zero if they are not connected), and $\eta_i$ accounts for a additive disturbances, including noise (typically considered white and Gaussian) and a potential forcing. 
Aggregating the phases and frequencies in a vector $\bm{X}_t \coloneqq (\bm{\theta}_t^\top, \bm{\omega}_t^\top)^\top$, we write Eq.~\eqref{eq:swing} as the stochastic differential equation, for $t\in[0,T]$,
\begin{align}\label{eq:dyn}
    d\bm{X}_t &= \bm{A}\bm{X}_tdt + \gamma\bm{e}_l{\rm Re}\left(e^{2\pi i(f t + \phi)}\right)dt + d\bm{W}_t, \\
    \bm{A} &= \begin{pmatrix}
        \bm{0} & \mathbb{I} \\
        \bm{M}^{-1}\bm{L} & -\bm{M}^{-1}\bm{D}
    \end{pmatrix}\, ,
\end{align}
where the matrices $\bm{M}$ and $\bm{D}$ are the diagnonal matrices of inertias and dampings respectively, $\bm{L}$ is the Laplacian matrix of the grid, weighted by the susceptances, $\gamma$, $f$, and $\phi$ are the amplitude, frequency and phase of the forcing respectively, $\mathbb{I}$ is the identity matrix of appropriate size, $\bm{e}_l$ is the canonical basis vector with nonzero $l$th component, and $\bm{W}_t$ is a Wiener process, accounting for unpredictable disturbances. Therefore, this model is a particular case of the dynamic model [Eq. (1) in the main text] considered in this work.

\section{Parameters of the studied synthetic test cases}
\label{app:test_cases_details}

\paragraph*{\bf Three-node test case in Fig.~1 (main text).} The dynamic state matrix $\bm{A}$ (see Eq.~(\ref{eq:dyn})) used for the example of Fig.~1 (main text) is given by:
\begin{align}\label{eq:3nodes}
    \bm{A} &= \begin{pmatrix}
        0 & 0 & 0 & 1 & 0 & 0 \\
        0 & 0 & 0 & 0 & 1 & 0 \\
        0 & 0 & 0 & 0 & 0 & 1 \\
        -\frac{417}{17} & \frac{40}{17} & -\frac{40}{17} & -\frac{1}{10} & 0 & 0 \\ 
        \frac{40}{17} & -\frac{21}{17} & \frac{4}{17} & 0 & -1 & 0 \\ 
        -\frac{40}{17} & \frac{4}{17} & -\frac{21}{17} &0 & 0 & -10 
    \end{pmatrix}\, .
\end{align}
It corresponds to the dynamics of three nodes, symmetrically coupled, with damping parameters varying over two orders of magnitude. Choosing these parameters and selecting a forcing frequency close to the imaginary part of the eigenvalues of matrix $A$, Eq.~(\ref{eq:3nodes}) allows one to have the largest amplitude of oscillation at a node that is not the source of the forcing and at a frequency that is not exactly the one of the forcing, but instead corresponding to the imaginary part of the selected system mode.\\

\paragraph*{\bf UK high-voltage network test case in Fig.~2 (main text).} The network topology used to generate the time series in the results of Fig.~2 is a model of the UK high-voltage transmission grids. It contains 120 nodes and 165 lines. Its adjacency list is given in Table~\ref{table1}.

\begin{table}
\begin{adjustbox}{width=0.52\columnwidth,center}
\begin{tabular}{lll|lll|lll|lll|lll|lll|lll}
1   & 2   & 1 & 21  & 120 & 1 & 42  & 118 & 1 & 68 & 67 & 1 & 77 & 78 & 1 & 95  & 97  & 1 & 115 & 116 & 1 \\
1   & 5   & 1 & 120 & 22  & 1 & 117 & 118 & 1 & 54 & 66 & 1 & 75 & 74 & 1 & 95  & 96  & 1 & 80  & 81  & 1 \\
2   & 3   & 1 & 22  & 26  & 1 & 117 & 35  & 1 & 66 & 65 & 1 & 72 & 76 & 1 & 97  & 96  & 1 & 16  & 20  & 1 \\
3   & 4   & 1 & 22  & 28  & 1 & 27  & 28  & 1 & 65 & 64 & 1 & 76 & 78 & 1 & 97  & 98  & 1 & 20  & 22  & 1 \\
3   & 11  & 1 & 9   & 24  & 1 & 28  & 29  & 1 & 64 & 63 & 1 & 78 & 79 & 1 & 94  & 107 & 1 & 35  & 36  & 1 \\
5   & 4   & 1 & 8   & 36  & 1 & 29  & 30  & 1 & 63 & 62 & 1 & 79 & 80 & 1 & 96  & 100 & 1 & 25  & 33  & 1 \\
4   & 6   & 1 & 7   & 42  & 1 & 29  & 31  & 1 & 62 & 61 & 1 & 78 & 80 & 1 & 99  & 100 & 1 & 25  & 26  & 1 \\
6   & 7   & 1 & 7   & 38  & 1 & 31  & 59  & 1 & 64 & 56 & 1 & 79 & 81 & 1 & 100 & 104 & 1 & 24  & 25  & 1 \\
7   & 8   & 1 & 38  & 39  & 1 & 117 & 49  & 1 & 56 & 57 & 1 & 81 & 89 & 1 & 100 & 101 & 1 & 24  & 34  & 1 \\
8   & 9   & 1 & 39  & 40  & 1  & 49  & 48  & 1 & 57 & 62 & 1 & 75 & 82 & 1 & 101 & 102 & 1 & 26  & 120 & 1 \\
11  & 10  & 1 & 38  & 40  & 1 & 48  & 46  & 1 & 57 & 58 & 1 & 82 & 83 & 1 & 102 & 104 & 1 & 46  & 118 & 1 \\
11  & 119 & 1 & 40  & 41  & 1  & 46  & 47  & 1 & 58 & 59 & 1 & 83 & 84 & 1 & 102 & 103 & 1 & 56  & 65  & 1 \\
119 & 10  & 1 & 41  & 37  & 1 & 46  & 45  & 1 & 59 & 60 & 1 & 84 & 85 & 1 & 104 & 105 & 1 & 68  & 69  & 1 \\
10  & 12  & 1 & 37  & 35  & 1 & 45  & 43  & 1 & 60 & 61 & 1 & 83 & 86 & 1 & 104 & 106 & 1 & 27  & 31  & 1 \\
10  & 9   & 1 & 35  & 34  & 1 & 45  & 49  & 1 & 60 & 74 & 1 & 86 & 87 & 1 & 107 & 108 & 1 & 93  & 94  & 1 \\
9   & 19  & 1 & 34  & 33  & 1 & 45  & 66  & 1 & 61 & 74 & 1 & 87 & 88 & 1 & 108 & 106 & 1 &     &     &   \\
12  & 13  & 1 & 33  & 32  & 1 & 49  & 50  & 1 & 73 & 74 & 1 & 87 & 90 & 1 & 107 & 109 & 1 &     &     &   \\
13  & 14  & 1 & 32  & 27  & 1  & 50  & 51  & 1 & 73 & 63 & 1 & 90 & 91 & 1 & 108 & 109 & 1 &     &     &   \\
14  & 15  & 1 & 26  & 27  & 1 & 50  & 54  & 1 & 73 & 72 & 1 & 89 & 95 & 1 & 109 & 112 & 1 &     &     &   \\
15  & 16  & 1 & 32  & 58  & 1 & 54  & 55  & 1 & 73 & 75 & 1 & 90 & 92 & 1 & 112 & 111 & 1 &     &     &   \\
16  & 17  & 1 & 33  & 57  & 1 & 54  & 52  & 1 & 72 & 71 & 1 & 91 & 92 & 1 & 110 & 111 & 1 &     &     &   \\
17  & 18  & 1 & 34  & 44  & 1 & 52  & 51  & 1 & 71 & 70 & 1 & 92 & 93 & 1 & 111 & 113 & 1 &     &     &   \\
17  & 19  & 1 & 43  & 44  & 1 & 52  & 53  & 1 & 70 & 69 & 1 & 92 & 94 & 1 & 114 & 113 & 1 &     &     &   \\
19  & 21  & 1 & 44  & 57  & 1 & 54  & 68  & 1 & 70 & 66 & 1 & 93 & 95 & 1 & 114 & 112 & 1 &     &     &   \\
19  & 23  & 1 & 42  & 43  & 1 & 54  & 67  & 1 & 69 & 77 & 1 & 94 & 95 & 1 & 114 & 115 & 1 &     &     &  
\end{tabular}
\end{adjustbox}
\caption{Adjacency list of the model of the UK high-voltage transmission grid. Each cell of the seven columns corresponds to line parameters: two indices that indicate two nodes at the end of the line, and the respective line capacity.}
\label{table1}
\end{table}

The time series are obtained by numerically solving Eq.~(1) in the main text, including a forcing at one bus.

\section{Derivation of the SALO and SALO-relaxed algorithms}
\label{app:derivation}

In this section, we provide the derivation of Eq.~(5) from the \emph{Methods} section of the main text. Starting with the negative likelihood expression (4), we have
\begin{align}
     L & \left(\bm{A},\gamma,l,k,\phi~|~\{\bm{X}_{t_j}\}_{j=1}^{N}\right)&\nonumber\\ 
     & = \frac{1}{N}\sum_{j=0}^{N-1}\left[\bm{\Delta}_{t_j} - \bm{A}\bm{X}_{t_j} - {\gamma}\bm{e_l} {\rm Re}\left(e^{2\pi i(\frac{k}{N}j + \phi)}\right)\right]^\top\left[\bm{\Delta}_{t_j} - \bm{A}\bm{X}_{t_j} - {\gamma}\bm{e_l}{\rm Re}\left(e^{2\pi i(\frac{k}{N}j + \phi)}\right)\right]\,\\
     \notag
     & = \frac{1}{N}\sum_{j=0}^{N-1} \Big[ \bm{\Delta}_{t_j}^\top\bm{\Delta}_{t_j} - 2\bm{\Delta}_{t_j}^\top \bm{A}\bm{X}_{t_j} + \bm{X}_{t_j}^\top\bm{A}^2\bm{X}_{t_j} + {\gamma}^2 \bm{e_l}^\top \bm{e_l}\,\left[{\rm Re}\left(e^{2\pi i(\frac{k}{N}j + \phi)}\right)\right]^2\\
     & \quad \quad \quad \quad \quad \quad \quad - 2{\gamma}\,\bm{e_l}^\top(\bm{\Delta}_{t_j} - \bm{A}\bm{X}_{t_j}){\rm Re}\left(e^{2\pi i(\frac{k}{N}j + \phi)}\right) \Big].
\end{align}
Define the discrete Fourier transforms of the data
\begin{align}
    \widetilde{\bm X}(k) &= \frac{1}{\sqrt{N}}\sum_{j=0}^{N-1} e^{2\pi i\frac{k}{N}j} \bm{X}_{t_j}\, , \\
    \widetilde{\bm \Delta}(k) &= \frac{1}{\sqrt{N}}\sum_{j=0}^{N-1} e^{2\pi i\frac{k}{N} j} \bm{\Delta}_{t_j}\, .
\end{align}
Then
\begin{align}
L\left(\bm{A},\gamma,l,k~|~\{\bm{X}_{t_j}\}_{j=1}^{N}\right) & = \frac{1}{N} \sum_{j=0}^{N-1} \left[ \bm{\Delta}_{t_j}^\top\bm{\Delta}_{t_j} - 2\bm{\Delta}_{t_j}^\top\bm{A}\bm{X}_{t_j} + \bm{X}_{t_j}^\top\bm{A}^2\bm{X}_{t_j} \right]
\label{eq:likelihood_app_term1}
\\
& + \frac{{\gamma}^2}{N} \sum_{j=0}^{N-1} \left[{\rm Re}\left(e^{2\pi i(\frac{k}{N}j + \phi)}\right)\right]^2\,
\label{eq:likelihood_app_term2}
\\
& - \frac{2{\gamma}}{\sqrt{N}}\,{\rm Re}\left(([\tilde{\bm \Delta}]_l - [\bm{A}\tilde{\bm{X}}]_l) e^{2\pi i \phi} \right)\,,
\label{eq:likelihood_app_term3}
 \end{align}
where $[{\bm Y}]_l$ is the $l$th component of a vector ${\bm Y}$.

Let us consider each of the terms in this expression separately. In \eqref{eq:likelihood_app_term1}, the term $\frac{1}{N}\sum_{j=0}^{N-1} \bm{\Delta}_{t_j}^\top\bm{\Delta}_{t_j}$ is independent of the variables being optimized, and hence can be dropped from the optimization. Using the definitions
\begin{align}
 \Sigma_0 = \frac{1}{N}\sum_{j=0}^{N-1}\bm{X}_{t_j}\bm{X}_{t_j}^\top\, , \quad \Sigma_1 = \frac{1}{N}\sum_{j=0}^{N-1}\bm{X}_{t_j}\bm{\Delta}_{t_j}^\top\, ,
\end{align}
the relevant terms in \eqref{eq:likelihood_app_term1} can be simply written as ${\rm Tr}(\bm{A}^\top\bm{A}\Sigma_0) - 2{\rm Tr}(\bm{A}\Sigma_1)$.

Regardless of the values of $\phi$ in \eqref{eq:likelihood_app_term2}, it can be shown that this term is simply equal to $\frac{1}{2}\gamma^2$. Indeed,
\begin{align}
    \frac{{\gamma}^2}{N} \sum_{j=0}^{N-1} \left[{\rm Re}\left(e^{2\pi i(\frac{k}{N}j + \phi)}\right)\right]^2 & = \frac{{\gamma}^2}{N} \sum_{j=0}^{N-1} \left[\frac{e^{2\pi i(\frac{k}{N}j + \phi)} + e^{- 2\pi i(\frac{k}{N}j + \phi)}}{2}\right]^2
    \\
    & = \frac{{\gamma}^2}{4N} \sum_{j=0}^{N-1} \left[ e^{4\pi i(\frac{k}{N}j + \phi)} + e^{- 4\pi i(\frac{k}{N}j + \phi)} + 2\right]
    \\
    & = \frac{{\gamma}^2}{2} + e^{4\pi i\phi} \frac{{\gamma}^2}{N} \sum_{j=0}^{N-1} e^{4\pi i \frac{k}{N}j} + e^{-4\pi i\phi} \frac{{\gamma}^2}{N} \sum_{j=0}^{N-1} e^{-4\pi i \frac{k}{N}j}
    \\
    & = \frac{{\gamma}^2}{2},
\end{align}
where in the last line we used the fact that $k$ is an integer.

Finally, notice that the last term \eqref{eq:likelihood_app_term3} is the only one that depends on the phase $\phi$, and the minimization over the phase $\phi$ can be performed independently of other variables. The minimum of \eqref{eq:likelihood_app_term3} over $\phi$ is given by $- \frac{2{\gamma}}{\sqrt{N}} \left| [\tilde{\bm \Delta}]_l - [\bm{A}\tilde{\bm{X}}]_l \right|$, where $\vert \cdot \vert$ is the modulus of a complex number. Further, we can write
\begin{align}
    - \frac{2{\gamma}}{\sqrt{N}} \left| [\tilde{\bm \Delta}]_l - [\bm{A}\tilde{\bm{X}}]_l \right| & = - \frac{2{\gamma}}{\sqrt{N}} \sqrt{\left([\tilde{\bm \Delta}]_l - [\bm{A}\tilde{\bm{X}}]_l)\right) \left([\tilde{\bm \Delta}]_l - [\bm{A}\tilde{\bm{X}}]_l\right)^\dagger}
    \\
    & = - \frac{2\gamma}{\sqrt{N}} \sqrt{{\rm Tr}\left(\bm{A}_{l,\cdot}^\top \bm{A}_{l,\cdot} F(k)\right) - 2f_l(k) \bm{A}_{l,\cdot} + g_l(k)},
\end{align}
where $M_{l,\cdot}$ denotes the $l$th row of the matrix, and we defined
\begin{align}
 F = {\rm Re}\left(\tilde{\bm X}\tilde{\bm X}^\dagger\right)\, , \quad 
 f_l = {\rm Re}\left([\tilde{\bm \Delta}]_l\tilde{\bm X}^\dagger\right)\, , \quad
 g_l &= \left|[\tilde{\bm \Delta}]_l\right|^2\, .
\end{align}

Therefore, using all the transformation above and defining a new objective function where minimization of $L\left(\bm{A},\gamma,l,k~|~\{\bm{X}_{t_j}\}_{j=1}^{N}\right)$ over $\phi$ has been performed,
\begin{equation}
    L\left(\bm{A},\gamma,l,k~|~\{\bm{X}_{t_j}\}_{j=1}^{N}\right) = \min_{\phi} L\left(\bm{A},\gamma,l,k,\phi~|~\{\bm{X}_{t_j}\}_{j=1}^{N}\right),
\end{equation}
we finally obtain the expression for the SALO objective function given in the \emph{Methods} section of the main text: 
\begin{align}
  L_{\rm SALO} & \left(\bm{A},\gamma,l,k ~|~\{\bm{X}_{t_j}\}_{j=0}^{N-1}\right)\nonumber
  \\
  &= {\rm Tr}(\bm{A}^\top\bm{A}\Sigma_0) - 2{\rm Tr}(\bm{A}\Sigma_1) + \frac{1}{2}\gamma^2 - \frac{2\gamma}{\sqrt{N}} \sqrt{{\rm Tr}\left(\bm{A}_{l,\cdot}^\top \bm{A}_{l,\cdot} F(k)\right) - 2f_l(k) \bm{A}_{l,\cdot} + g_l(k)}\, .
  \label{eq:SALO_app}
\end{align}

In order to obtain the SALO-relaxed version of the algorithm, we notice that the objective \eqref{eq:SALO_app} can be rewritten in an equivalent form, where the candidate source node $l$ and the amplitude $\gamma$ are encoded as the $l$th component of the one-hot vector $\bm{\gamma}$ that spans all nodes in the system:
\begin{align}
  L_{\rm SALO} & \left(\bm{A},\bm{\gamma},k~|~\{\bm{X}_{t_j}\}_{j=0}^{N-1}\right)\nonumber
  \\
  & = {\rm Tr}(\bm{A}^\top\bm{A}\Sigma_0) - 2{\rm Tr}(\bm{A}\Sigma_1) + \sum_{l=1}^n \left[\frac{\gamma_l^2}{2} - \frac{2\gamma_l}{\sqrt{N}} \sqrt{{\rm Tr}\left(\bm{A}_{l,\cdot}^\top\bm{A}_{l,\cdot}F(k)\right) - f_l(k)\bm{A}_{l,\cdot} + g_l(k)} \right]\, ,\label{eq:SALO_app_equiv}\\
  & \text{s.t.} \quad \Vert \bm{\gamma} \Vert_0 = 1.\nonumber
\end{align}
A spatially relaxed version of the algorithm over the variable $\gamma$ is obtained by dropping the one-hot-encoding constraint $\Vert \bm{\gamma} \Vert_0 = 1$, a version that we refer to as the SALO-relaxed algorithm: 
\begin{align}
 L_{\rm SALOr} & \left(\bm{A},\bm{\gamma},k~|~\{\bm{X}_{t_j}\}_{j=1}^N\right)\nonumber
 \\
 &= {\rm Tr}(\bm{A}^\top\bm{A}\Sigma_0) - 2{\rm Tr}(\bm{A}\Sigma_1) + \sum_{l=1}^n \left[\frac{\gamma_l^2}{2} - \frac{2\gamma_l}{\sqrt{N}} \sqrt{{\rm Tr}\left(\bm{A}_{l,\cdot}^\top\bm{A}_{l,\cdot}F(k)\right) - f_l(k)\bm{A}_{l,\cdot} + g_l(k)} \right]\,.
 \label{eq:SALO_relaxed_app}
\end{align}
In this version of the algorithm, the source location is determined from the indices of the largest components of the vector $\bm{\gamma}$.

\section{Challenges for the optimization of the continuous likelihood function}
\label{app:challenges_likelihood}

\begin{figure}[!htb]
    \centering
    \includegraphics[width=0.5\columnwidth]{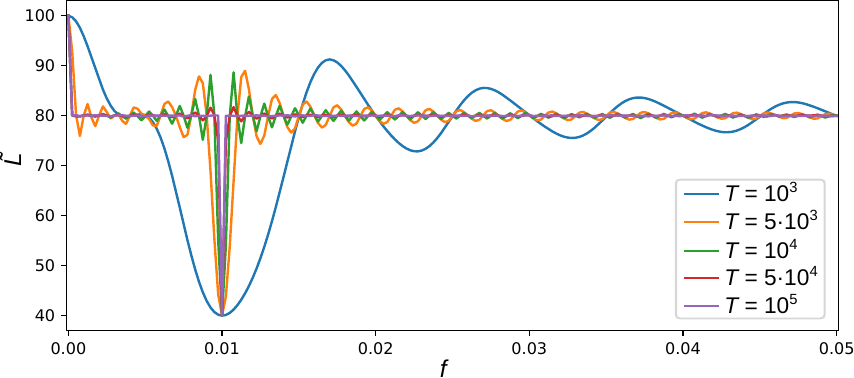}
    \caption{Objective function \eqref{eq:L_continuous} for a small test problem plotted as a one-dimensional function of the forcing frequency $f$ for different observation windows $T$. For simplicity, all other parameters of the model ($\bm{A}$, $\gamma$, $l$, and $\phi$) are set to their ground-truth values used to generate the data. With increasing $T$, the objective function becomes more and more challenging to optimize, exhibiting an oscillatory behavior with a narrow peak at the correct frequency value.}
    \label{fig:obj_w}
\end{figure}

The length and the sampling rate of the available time series impose a natural limit to the frequency resolution, which sets a limit on a precision with which the forcing frequency $f$ can be identified. This idea lies at the heart of the discretization in the frequency domain explained in the \emph{Methods} section of the main text. Both SALO and SALO-relaxed methods are then evaluated for each discrete value of possible frequencies $k$. To provide an additional insight into our approach, in this section we illustrate why a direct optimization over the continuous frequency $f$ results in a hard-to-solve non-convex optimization problem. Consider a direct optimization over a continuous variable $f$ of the negative log-likelihood
\begin{align}\label{eq:L_continuous}
 \widetilde{L}\left(\bm{A},\gamma,l,f,\phi~|~\{\bm{X}_{t_j}\}_{j=1}^{N}\right) &= \frac{1}{N}\sum_{j=0}^{N-1}\left\|\bm{\Delta}_{t_j} - \bm{A}\bm{X}_{t_j} - \gamma\bm{e}_l\cos(2\pi i(f t_j + \phi))\right\|^2
\end{align}
from $N$ measurements $\{\bm{X}_{j\tau}\}_{j=0,\ldots,N-1}$ taken at times $t_j=j\tau$ over the observation window of length $T$, so that the interval $\tau=T/N$. For simplicity, let us fix variables $\bm{A}$, $\gamma$, $l$, and $\phi$ to their ground-truth values. In this case, we can plot the objective function \eqref{eq:L_continuous} for a test problem as a one-dimensional function of $f$, see Figure~\ref{fig:obj_w}. We see that with the increasing observation window $T$, the objective function is oscillating around a certain constant value for most values of $f$, with a sharp and narrow peak at the correct frequency value.

\begin{figure}[!htb]
 \centering
 \includegraphics[width=0.93\textwidth]{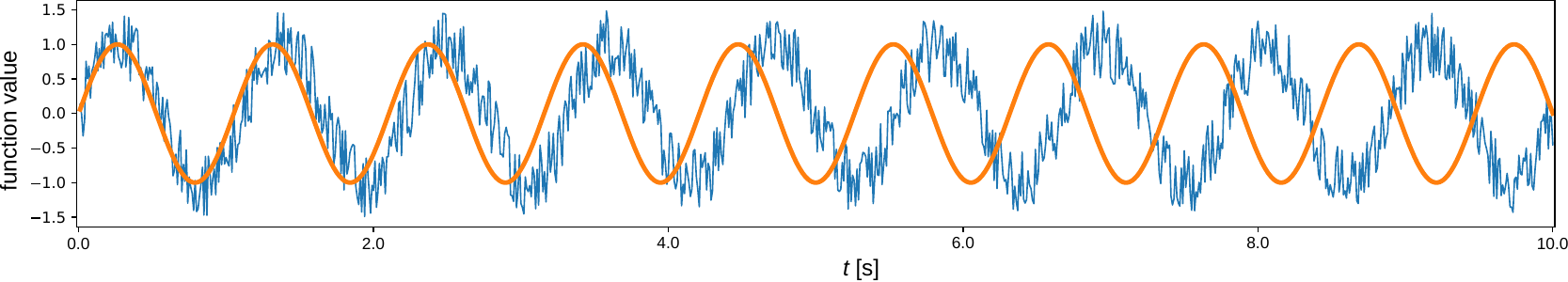}
 \caption{Example of the divergence between a time series with the period $\nu=0.95$ $[{\rm s}^{-1}]$ and a sinusoidal fit with a mismatching frequency $f=0.90$ $[{\rm s}^{-1}]$.}
 \label{fig:wrong_freq}
\end{figure}

It is possible to understand this behavior using a toy example presented in Figure~\ref{fig:wrong_freq}. When parameters $\bm{A}$, $\gamma$, $l$, and $\phi$ are fixed to their ground-truth values, the optimization problem in \eqref{eq:L_continuous} becomes essentially equivalent to fitting of a noisy periodic function $\cos(\nu t) + \xi_t$, where $\xi_t$ is a white Gaussian noise, with a function $\cos(f t)$. For any value of $f$ which is not equal to $\nu$ and a sufficiently long time series, a small deviation of $f$ from $\nu$ makes the fit eventually diverge to the point where the two curves appear in counterphase. This effect is responsible for the form of the objective function presented in the Figure~\ref{fig:obj_w}. This type of landscapes with a large number of local minima and a single narrow global minimum is particularly challenging for optimization solvers. This challenge is alleviated in our approach by switching to the discrete frequencies $k$ and turning the minimization into a mixed-integer optimization problem.     

\section{Performance of the SALO-relaxed algorithm}
\label{app:l1-relaxation}

The SALO-relaxed algorithm has an improved computational time compared to the SALO algorithm given that it does not need to be run for all possible candidate sources in the network. In this section, we empirically benchmark the SALO-relaxed algorithm against the SALO method. For all the test cases used throughout the main text, we observed that the SALO-relaxed algorithm points to the same source, consistently with SALO. Here, we use additional test cases to illustrate that while the SALO-relaxed algorithm provides an improvement in computational time, it does not sacrifice the accuracy of the source localization.

\begin{figure}[!htb]
 \centering
 \includegraphics[width=0.84\textwidth]{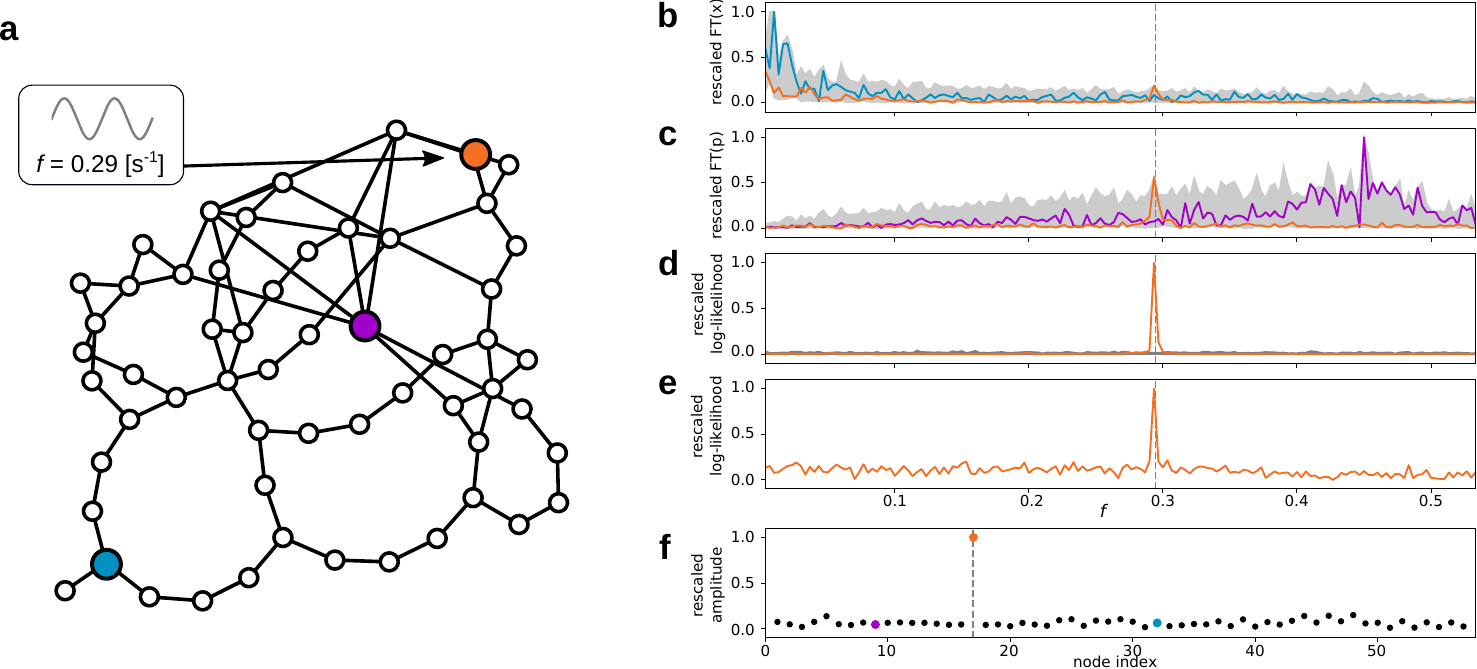}
 \caption{{\bf (a)} Synthetic test case with the IEEE-57 test case topology, designed to reproduce the resonance conditions where forced oscillations interact with the system modes. The forcing at the orange node results in a highest response in Fourier spectra at the opposite side of the network, as shown for the Fourier components of the {\bf (b)} generalized state and of the {\bf (c)} generalized momentum. {\bf (d)} SALO algorithm confidently identifies the correct forcing frequency and source without any knowledge of the system topology or parameters. The SALO-relaxed version of the method is equally capable to unambiguously identify both {\bf (e)} the frequency and {\bf (f)} the source of forced oscillations under these challenging conditions.
 }
 \label{fig:ieee57}
\end{figure}

In Figure~\ref{fig:ieee57}, we compare the outcome of both methods on the network structure extracted from the standard IEEE 57-bus test case~\cite{IEEE57}. Similarly to the test case presented in the Figure~2 in the main text, we modified the parameters of the network to produce the challenging resonance conditions, where the largest response seen in the Fourier spectrum appears on nodes that are far away from the source. Importantly, under these challenging conditions, SALO-relaxed method agrees with the SALO algorithm, and confidently points both to the correct source of forced oscillations, as well as to the ground-truth forcing frequency.  

\begin{figure}[!htb]
    \centering
    \includegraphics[width=\textwidth]{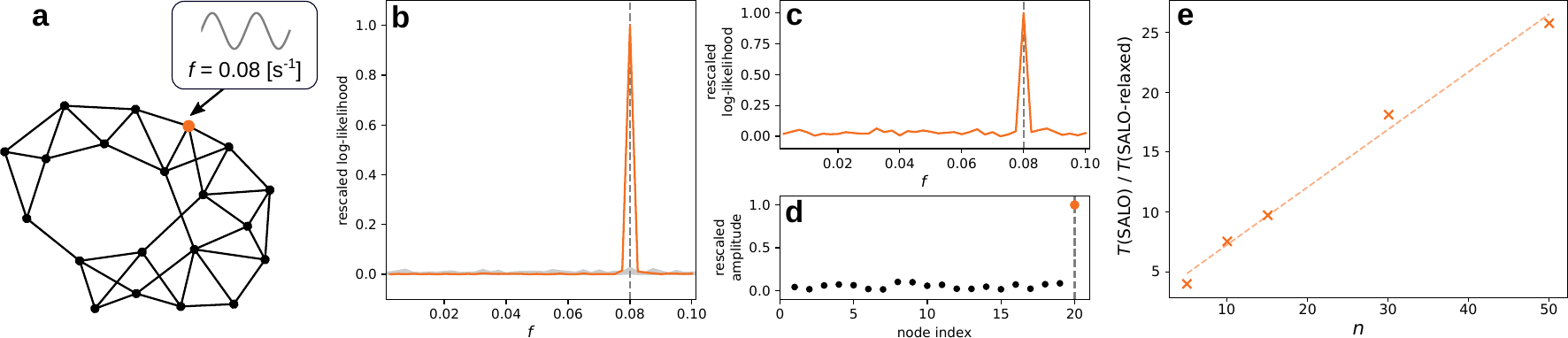} 
    \caption{{\bf (a)} A random instance of a Watts-Strogatz network with $20$ nodes and one source of forced oscillations (shown in orange). {\bf (b)} Rescaled log-likelihood for the SALO algorithm on a random network of $20$ nodes, which correctly identifies the oscillation source and the frequency. Similarly, the SALO-relaxed method perfectly identifies both {\bf (c)} the frequency and {\bf (d)} the location of the source in this network. {\bf (e)} Ratio of SALO and SALO-relaxed run-times without parallelization over candidate sources $l$ as a function of the system size $n$\,. Dashed line shows a linear fit of the data points.}
    \label{fig:timing}
\end{figure}

In Figure~\ref{fig:timing}, we run a scaling experiment to compare the computational speed of SALO-relaxed and SALO methods, without using parallelized computation for the latter one. For this purpose, we use random Watts-Strogatz networks of increasing sizes. At every network instance, the source and frequency of forced oscillations is correctly identified by both algorithms, however, SALO-relaxed algorithm enjoys a computational advantage over SALO that scales linearly with the system size.

We provide further results of testing of the SALO-relaxed in the situation where some of the modeling assumptions are violated in the next section \ref{app:real-data-additional}.

\section{Additional results on the real PMU data}
\label{app:real-data-additional}

\begin{figure}[!htb]
    \centering
    \includegraphics[width=\textwidth]{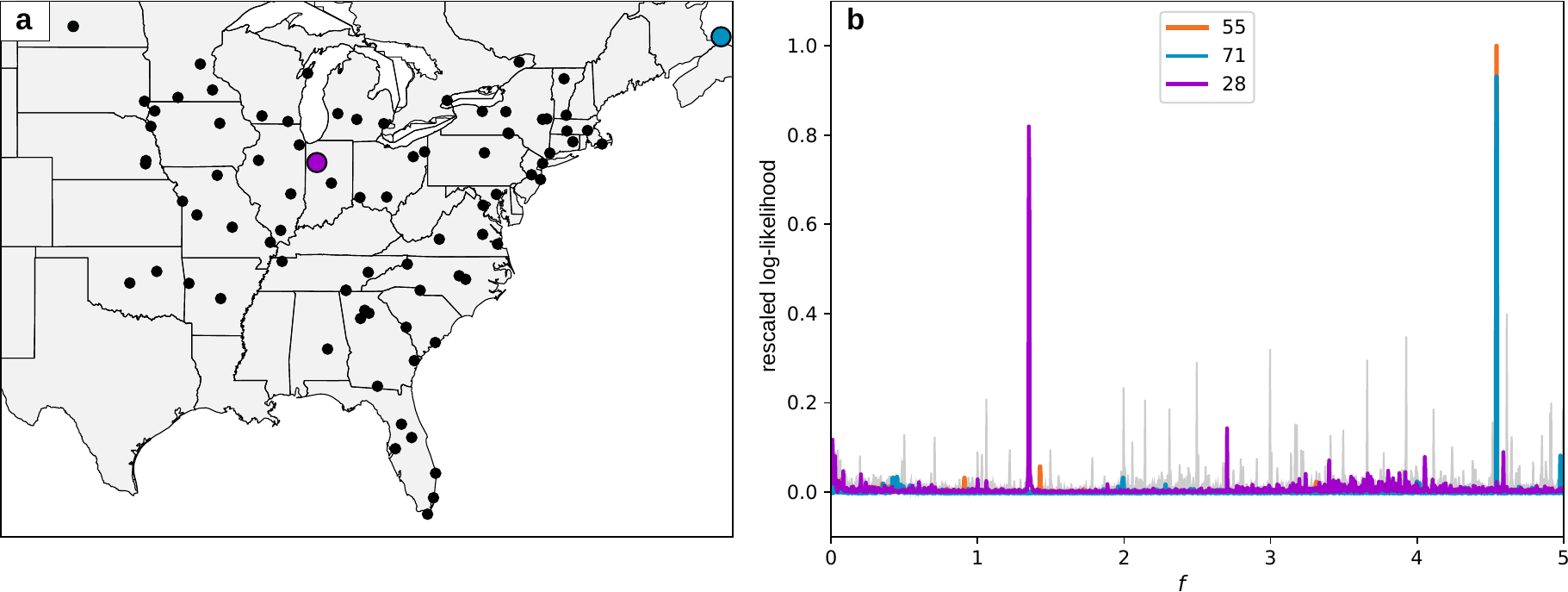}
    \caption{{\bf (a)} Known locations of FNET/GridEye measurement devices that recorded the data under an oscillatory event in the U.S. Eastern Interconnection area \cite{UTKdata}, along with the location of the most likely sources identified by SALO. {\bf (b)} Rescaled log-likelihood scores peaking at two different frequencies for the most likely sources of oscillations. The location of the orange node is not available.}
    \label{fig:fig4sup_utk}
\end{figure}

In the Figure~3 of the main text, we showed the results produced by the SALO method in the case of multiple sources injecting oscillations at different frequencies, potentially with sources lying outside of the observed system. In this section, we apply the SALO method to real data, which shows a combination of features observed in Figure~3 of the main text. We use the data collected under an oscillatory event in the U.S. Eastern Interconnection from FNET/GridEye, a GPS-synchronized wide-area frequency measurement network \cite{UTKdata}. We have analyzed the subset of the time series without missing data. The ground-truth location of the sources of forced oscillations is not known in the considered data set. The network topology or parameters are also not known, which calls for the application of the SALO method. The known locations of a subset of measurement devices is depicted in Figure~\ref{fig:fig4sup_utk}~(a). The analysis of the time series with our SALO method points to the most likely sources of forced oscillation, highlighted with the orange, blue, and purple colors in the rescaled log-likelihood scores in Figure~\ref{fig:fig4sup_utk}~(b). A precise location of the node corresponding to the likelihood score highlighted in orange is not available, however the other likely sources are shown with the blue and purple colors in Figure~\ref{fig:fig4sup_utk}~(a).

\begin{figure}[!htb]
    \centering
    \includegraphics[width=\textwidth]{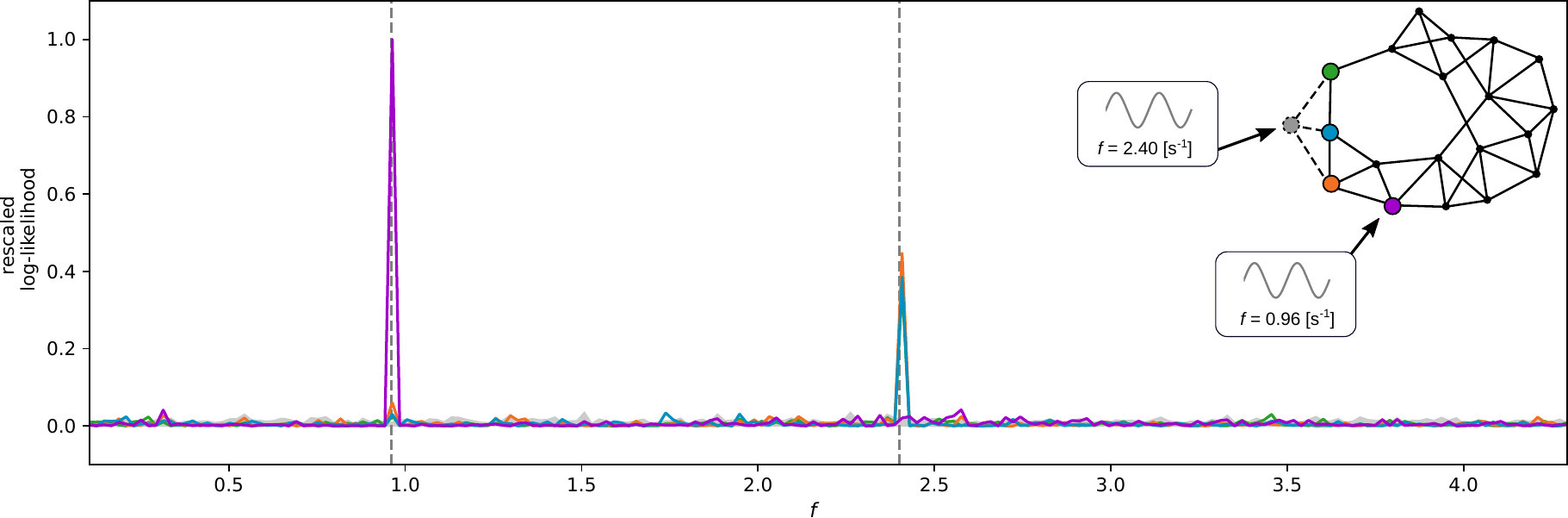}
    \caption{Rescaled log-likelihood score obtained by SALO for the scenario with one observed and one hidden sources of forced oscillations. The algorithm correctly identifies the location and forcing frequency for the visible source, and points to the neighbors of the hidden source inside the visible system for the detected oscillation at the other forcing frequency. 
    As before, the envelope of scores for non-highlighted nodes is shown in gray.}
    \label{fig:fig4sup_l0}
\end{figure}

The features shown in the log-likelihood scores are reminiscent of the ones observed in the synthetic study Figure~3 of the main text. In particular, both the orange and the blue nodes indicate the presence of oscillations at the exact same frequency $f=4.54$ $[{\rm s}^{-1}$]. This feature is similar to the response of the algorithm observed in the case where the source is located outside of the observed system, see Figure~3~(a) in the main text. Although the geographical location of the orange node is not available, the blue node is indeed located at the edge of the observed area, which makes this scenario plausible. The analysis with SALO also points to a presence of the second source of oscillations at a different frequency $f=1.35$ $[{\rm s}^{-1}]$, similarly to the algorithm response observed in a synthetic case with two oscillation sources shown in Figure~3~(b) of the main text. The respective source node found by SALO is located inside the bulk of the system, as shown in purple in Figure~\ref{fig:fig4sup_utk}~(a). In Figure~\ref{fig:fig4sup_l0}, we use a synthetic network example to verify that the combination of these features -- two sources of oscillations, inside and outside the observed system -- indeed leads to a picture corresponding to a combination of features observed in Figures 3~(a) and 3~(b) of the main text. Interestingly, the nature of the leading peaks in the rescaled log-likelihood observed in Figure~\ref{fig:fig4sup_l0} shows a lot of resemblance with the results produced by the SALO algorithm in Figure~\ref{fig:fig4sup_utk}~(b).     

\begin{figure}[!htb]
    \centering
    \includegraphics[width=\textwidth]{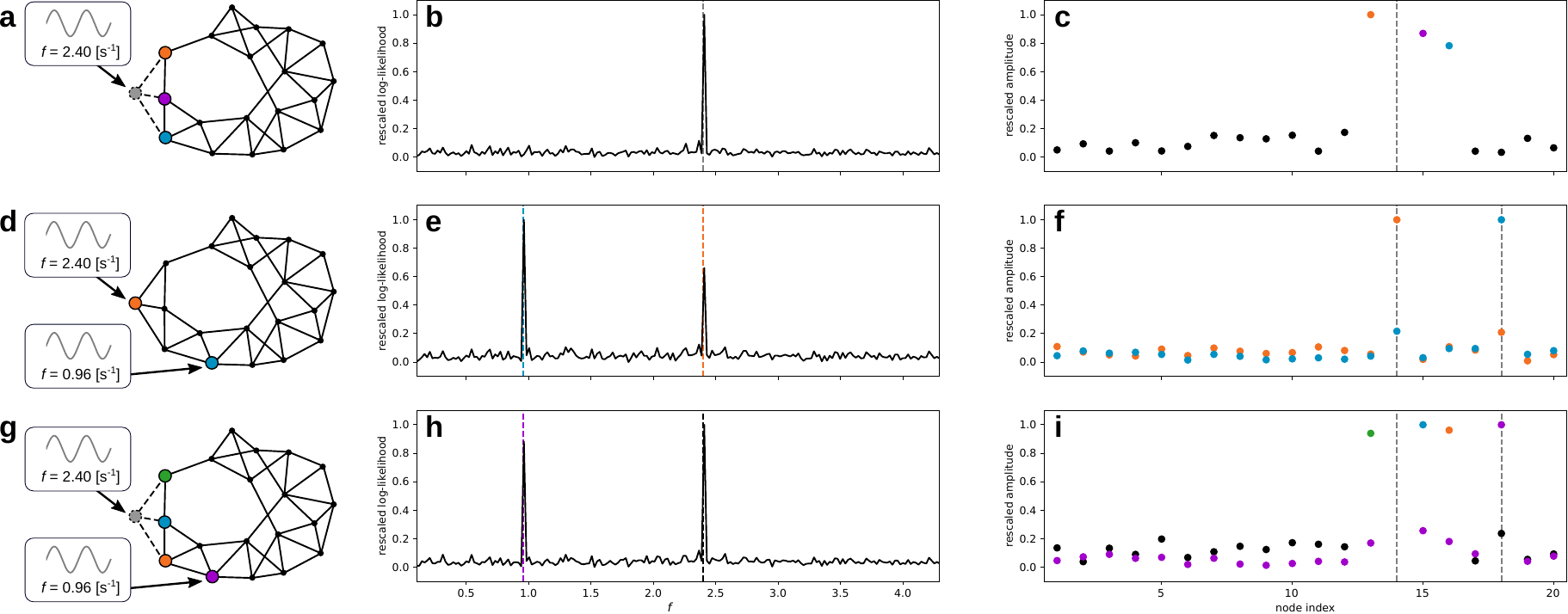}
    \caption{Rescaled log-likelihood and amplitude vector inferred using the SALO-relaxed version of the algorithm for three scenarios of model mismatch, illustrated on a network with $20$ nodes. In the top panel, forced oscillation is injected at the unobserved gray node {\bf (a)}. SALO-relaxed identifies the correct forcing frequency {\bf (b)} and points to the immediate neighbors {\bf (c)} of the gray node as the most likely sources. The center panel studies the situation where two independent forcings are present at different nodes {\bf (e)}. For two frequency peaks identified in the log-likelihood score {\bf (f)}, we show two respective amplitude vectors, with the largest components pointing to the correct sources. The bottom panel shows the combination of the previous two scenarios, with two sources, one observed (purple) and one unobserved (gray) {\bf (g)}. Both forcing frequencies {\bf (h)} and the respective locations from the largest components of the amplitude vector {\bf (i)} are correctly identified. All results are consistent with the outcome produced by SALO, see Figure~3~(a),(b) in the main text, and Figure~\ref{fig:fig4sup_l0}.}
    \label{fig:fig4sup_l1}
\end{figure}

Finally, we test the performance of the SALO-relaxed algorithm under the scenarios of violated modeling assumptions studied in Figures 3~(a) and 3~(b) of the main text, as well as for the combination of hidden and multiple sources examplified in Figure~\ref{fig:fig4sup_l0}. The results of these tests are reported in Figure~\ref{fig:fig4sup_l1}. In all three cases, the outcome produced by the SALO-relaxed algorithm is in complete agreement with the results shown by SALO. 

\section{Use of prior information on the system parameters}\label{informed}

\begin{figure}[!htb]
    \centering
    \includegraphics[width=\textwidth]{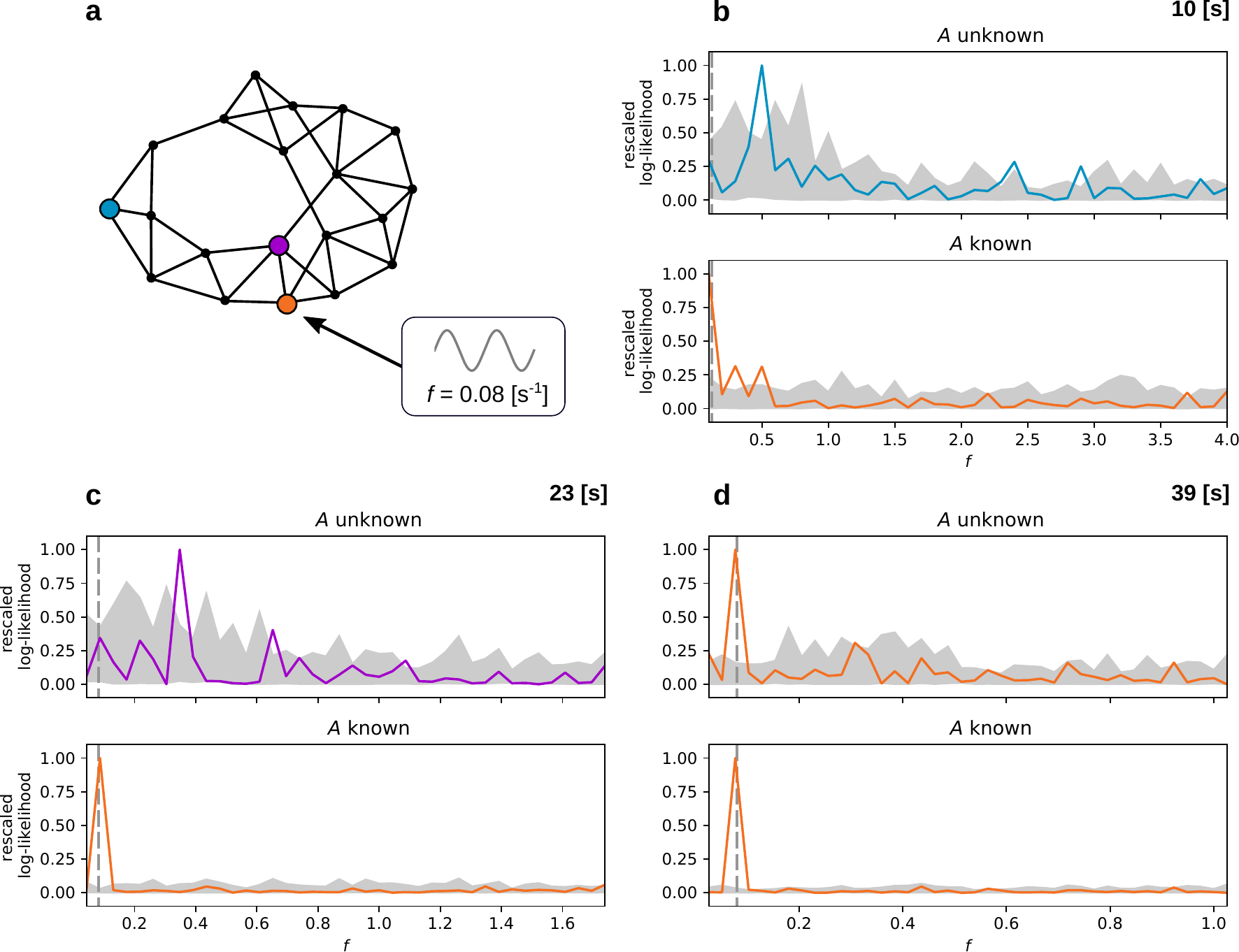}
    \caption{Rescaled log-likelihood score obtained by the SALO algorithm from time series data with three different lengths, illustrated on a network with $20$ nodes {\bf (a)} with the forced oscillation injected at node 20 at a frequency about $0.08$ [${\rm s}^{-1}$]. For shorter time series with length of {\bf (b)} 10 and {\bf (c)} 23 [s], the system-informed version of SALO algorithm is capable of correctly identifying the forcing, whereas the system-agnostic version requires more data since it simulataneous needs to estimate the parameter matrix $A$. The time series with length of {\bf (d)} 39 [s] are enough for both the system-agnostic and system-informed versions of the method to recover both location and frequency of the forced oscillation.}
    \label{figlast}
\end{figure}

\begin{figure}[!htb]
    \centering
    \includegraphics[width=\textwidth]{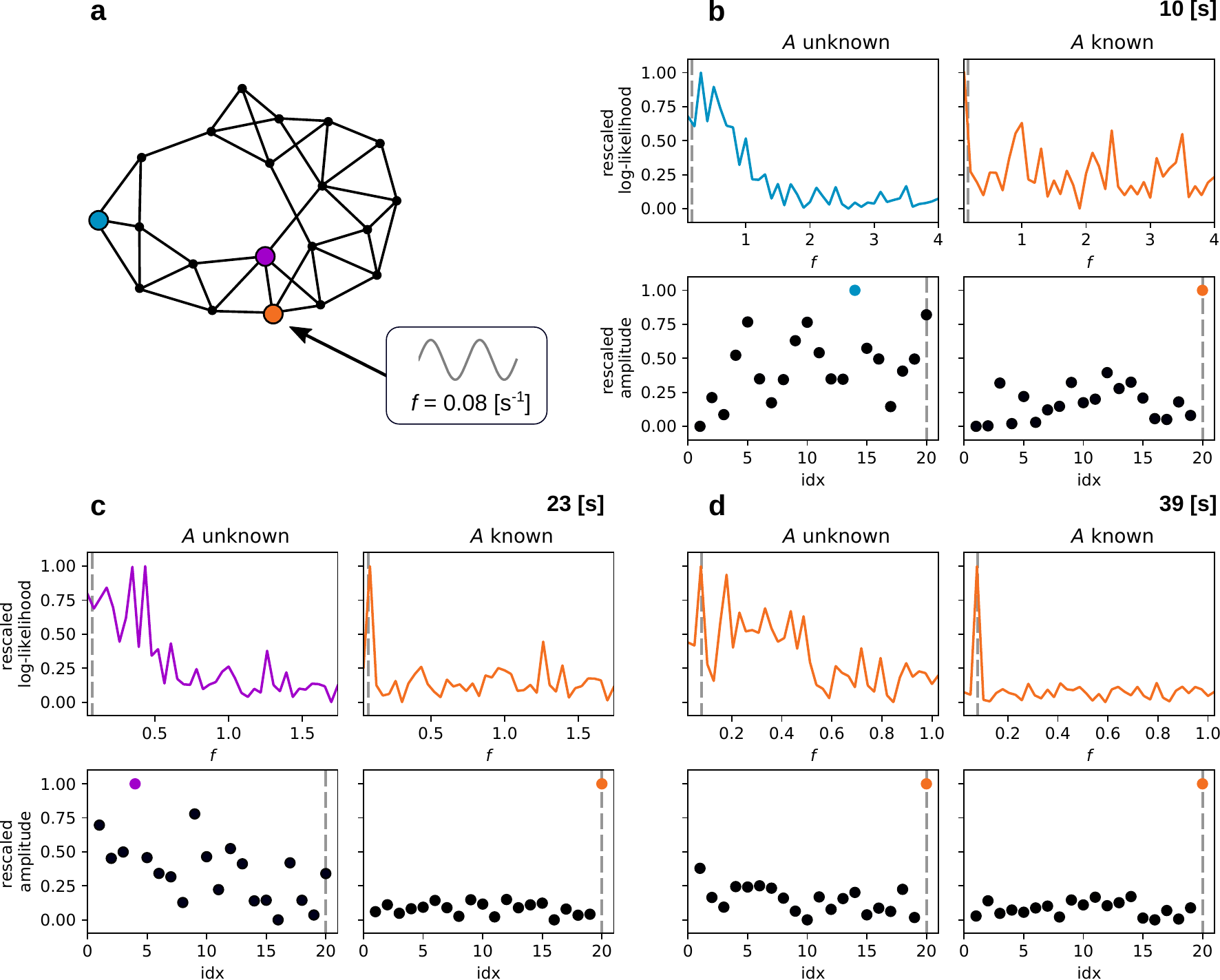}
    \caption{Rescaled log-likelihood score and location amplitude vector obtained by the SALO-relaxed algorithm from time series data with three different lengths, illustrated on a network with $20$ nodes {\bf (a)} with the forced oscillation injected at node 20 at a frequency about $0.08$ [${\rm s}^{-1}$]. For shorter time series with length of {\bf (b)} 10 and {\bf (c)} 23 [s], the system-informed version of SALO-relaxed algorithm is capable of correctly identifying the forcing, whereas the system-agnostic version requires more data since it simulataneous needs to estimate the parameter matrix $A$. The time series with length of {\bf (d)} 39 [s] are enough for both the system-agnostic and system-informed versions of the method to recover both location and frequency of the forced oscillation.}
    \label{figlast2}
\end{figure}

So far in the paper, we have considered the most challenging case where no information on the system parameters is assumed. In this section, we study the performance of SALO and SALO-relaxed methods in the scenario where the system topology and parameters, i.e.,  elements of the matrix $\bm A$ in Eq.~(\ref{eq:dyn}), are known. We compare the system-agnostic version of SALO with the system-informed one in Figures.~\ref{figlast} and \ref{figlast2}. For both versions of the algorithm, inclusion of prior information enables the correct identification of the source and the frequency of the forcing with significantly less data. We found that using time series of length below 250 time steps, the agnostic version was not able to correctly identify the forcing, whereas the system-informed version provided a correct recovery using time series with up to 100 time steps in length. These experiments highlight the reduction of the required data for a successful localization of the forced oscillations when the prior information on the system is available.

\bibliography{bibliography}

\end{document}